# Conceptual design of an innovative UVC-LED air-cleaner to reduce airborne pathogen transmission

Saket Kapse[2], Dena Rahman[1], Eldad J Avital[1*], Taylor Smith[1], Lidia Cantero-Garcia[1], Maham Sandhu[1], Rishav Raj[3], Fariborz Motallebi[1], Abdus Samad[3], Nithya Venkatesan[2], Clive B Beggs[4]

[1]School of Engineering and Materials Science, Queen Mary University of London, London UK
[2]School of Electrical Engineering, Vellore Institute of Technology, Chennai, Tamil Nadu, India
[3]Department of Ocean Engineering, Indian Institute of Technology, Madras, Tamil Nadu, India.
[4]Carnegie school of sport, Leeds Beckett University, Leeds UK
*Corresponding author: e.avital@qmul.ac.uk

**Abstract**

A conceptual design of a novel UVC-LED air-cleaner is presented as part of an international educational-research study. The main components are a dust-filter assembly, a UVC chamber and a fan. The dust-filter aims to suppress dust accumulation that will hamper the UVC chamber operation. The innovation is in the UVC chamber that includes a novel turbulence-generating grid to enhance air mixing in the chamber and a novel LEDs layout to achieve sufficient kill of the SARS-CoV-2 virus and TB bacterium aerosols with a reasonable power consumption. Both diseases have hit hard low to medium income countries and this study is part of an effort to offer non-pharmaceutical solutions to mitigate the air-transmission of such diseases. Low to high fidelity methods of computational fluid dynamics and UVC ray method are used to show that the design can provide a kill above 97% for Covid and TB, and above 92% for influenza-A. This is at a flow rate of 100 l/s, power consumption of less than 300W and a device size that is both portable and may also fit into ventilation ducts. Research and educational methodologies are discussed, along with analysis of the inexpensive dust-filter performance and the irradiation and flow fields.

# 1. Introduction

*1.1 Motivation and aims*

The Covid-19 pandemic has highlighted the need to control indoor air quality and in particular the need to reduce the concentration of aerosolized pathogens in room air. Respiratory pathogens can be transmitted directly by droplets emitted by an infected person, ballistically hitting the face (or food) of another person [1], in which case social distancing and use of screens can significantly reduce the transmission [2]. However, viral pathogens can also be transmitted by the airborne route. Small droplets (< 50 microns) and droplet nucelli (<20 microns) linger in the air [1, 2, 3], and potentially infect people at longer distances. Poorly maintained mechanical ventilation and air-conditioning systems with re-circulation can also spread airborne diseases [4].

Despite initial reluctance to recognize the Covid-19 as an airborne transmitted disease, strong evidence supports the airborne route theory [5-6]. It joins other infamous airborne diseases like tuberculosis (TB), measles, and influenza [7]. Therefore, effective room ventilation is important to reduce the pathogen load in the air [1, 2]. However, this is not always possible because: inclement weather conditions might prevent the opening of windows or the use of full 'fresh air' mechanical ventilation, or simply the building may have been constructed with poor ventilation options.

In such scenarios, an air-cleaning device called the high-efficiency particulate air (HEPA) filter helps, The air passes through the HEPA filter mounted inside a heating, ventilating and air conditioning (HVAC) system [2]. However, such filters require regular maintenance and replacement. Also, dust and other particulate matter collected on the HEPA filter create high resistance to airflow causing an increased energy consumption. Hence, the alternative technologies such as UVC light and/or air ionization, with self-cleaning filter technologies are evolved.

A UVC based air disinfection technology utilizes UV light of wavelength 220-280 nm to damage the genetic material (DNA or RNA) of pathogens (viruses or bacteria); and inactivates and prevents them from causing an infection [7]. Such technology has been implemented since the 1930s to mitigate the spread of TB [8], with open UVC lamps radiating light above people's heads of room occupants (i.e., upper room UVC air disinfection) generally the favoured configuration [2].

Recently, it has been suggested that upper-room UVC air disinfection might be effective at mitigating the spread of the SARS-CoV-2 virus in aerosols, e.g. [7,9]. The SARS-CoV-2 virus appears to be particularly vulnerable to damage from UVC light when in aerosols [7]. However, because the ir-radiation field is open the UVC lamps must be shielded from downwards view by baffles which reduces the intensity of the upper-room UVC field. Furthermore, care must be taken to prevent UVC light reflected into the lower room space to keep people within safety limits of the light intensity [8]. Such installation also requires ceiling fans to promote good mixing of air between the lower and the upper zones, so the residence time of the aerosol droplets in the upper-room irradiation zone is maximized. Hence, upper-room UVC systems are more suitable for large public indoor places with high ceilings such as schools, hospitals, and transport hubs where the installation costs can be justified.

Air disinfection can also be achieved by mounting UVC lamps inside a 'box' with a fan, which allows the technology to act as a stand-alone air cleaner that can be safely mounted in a room, or inside a duct of a mechanical ventilation system [10]. The design is similar to a HEPA filter, but a UVC chamber replaces a HEPA filter in this design. UVC air cleaner is a compact air cleaner which avoids damaging eyes and skin [11,12] as it is shielded inside the box.

However, such devices also introduce technological challenges, as it is difficult to achieve a sufficient residence time for the pathogens to be inactivated inside the UVC chamber, while still maintaining a reasonable flow rate at an acceptable financial cost and power consumption. This major technological challenge is not trivial, because it has the potential to undermine the effectiveness of such devices when utilised as an infection control intervention. Unfortunately, many wrongly assume that because a device can achieve a 99.9% disinfection rate based on a 'single-pass' microbiological test, that it will protect room occupants from acquiring airborne infection. However, such claims can be extremely misleading, because they relate purely to the air that passes through the UV device and not to the overall effect that the unit has on the room space. This is largely dictated by the volume flow rate at which the disinfected air is supplied to the space [13], as highlighted in Eq (1), showing the steady-state concentration achieved by an air cleaning device in a fully mixed room space;

$$C = \frac{q}{\dot{v}_r + \dot{v}_{ac}\eta} \quad . \tag{1}$$

C is the contaminant concentration in the room space under steady-state conditions (virions/m$^3$), q is the steady-state room contamination rate (virions/s), $\dot{v}_r$ is the room ventilation flow-rate (m$^3$/s), $\dot{v}_{ac}$ is the air flow rate through the UVC air cleaner (m$^3$/s), and η is the single-pass efficiency of the air disinfection device expressed as a fraction.

From Eq (1) it is seen that unless an air cleaning device can deliver a relatively high air flow rate, it is unlikely to have a major impact on the transmission of COVID-19 or other airborne diseases within a room space, even if all the air passing through the unit is 100% disinfected. Indeed, from an infection control standpoint, it is much better to install a less efficient device which delivers a large quantity of 'disinfected' air, rather than having a smaller device which delivers very clean air, but at a very low supply volume flow rate. So, the technical challenge is one of supplying as much clean air as possible to the room space, while still maintaining a relatively high level of disinfection and not consuming large amounts of energy. It is this challenge that we address in this study, where we describe the development of a novel 'UVC in a box' air-cleaner.

Light-emitting diodes (LEDs) that emit UVC light at a wavelength of 279 nm have been developed recently. They do not ionize the air and therefore produce no ozone, while potentially being more efficient in terms of power being radiated into short distances as compared to conventional low-pressure lamps at 254 nm. The LEDs are compact and do not use of mercury in their production [14]. Hence, this technology was chosen for our air-cleaning device for reasonable power consumption and cost suitable for low and medium-income countries. Hence, a compromise between delivering the maximum amount of disinfected air and ensuring that the single-pass efficiency of the device has been made while maintaining low energy consumption. Our aim is to design a device with a flow rate of 100 l/s and 95% kill of Covid-19 virus and TB bacterium. This provides an equivalent ventilation rate of six air changes per hour (ACH) in a large living room or mid-size office room of 6×4×2.5 $m^3$, which is the WHO's minimum requirement for a health setting where no aerosol-generating procedure occurs [15]. It can also support up to ten people with 10 l/s per person equivalent ventilation, which follows the WHO's recommendation for non-residential buildings [15].

To achieve aims, the design requires an efficient irradiation chamber to deliver the necessary UVC dose to inactivate the pathogens. This leads to the requirement of an optimised LED layout together with an innovative aerodynamic design which ensures that the aerosol droplets are sufficiently exposed to the UVC field with no dead zones giving escape route to the droplets without irradiation dose.

*1.2 The conceptual design*

The study ran as a one-year joint project between three universities, Queen Mary University of London -UK, IIT Madras and VIT Chennai- India with the support of the UK's Royal Academy of Engineering. The concept of the air-cleaner is illustrated in Fig 1, where an axial fan draws air through an irradiation chamber comprising four opposite facing banks of UVC-LEDs emitting light at 279 nm mounted in two cells. Keys to the design are: (i) A novel turbulence generating grid that mixes the air in the irradiation chamber, thus ensuring that the exposure time is maximised and that any pathogens which enter the chamber are evenly irradiated (ii) Smart layout of the LEDs to reduce power consumption and manufacturing costs, while achieving the required inactivation rate of pathogens.

In dusty environment, dust accumulation in the device can impair the LEDs' efficiency. A low-cost coarse dust filter is mounted in front of the UVC chamber to capture particles larger than 10 microns diameter. Most infectious droplets are thought to be in the region 2 to 20 microns [16], thus the filter may become clogged and contaminated over time, posing an infection risk. To facilitate the safe replacement of the filter, an isolating 'cassette' assembly can be used with the dust-filter residing between the two sets of louvres (as illustrated in Fig 2). When the cassette is slotted into the air cleaning device, the louvres are to be opened, and when it is time to replace the dust-filter, the louvres are closed and the 'cassette' assembly can be removed for cleaning (disinfection) and safe disposal of the dust-filter.

Having passed through the dust filter, the air passes through the turbulence-generating grids before entering the UVC chamber and then going into the suction fan. The indicative axial lengths of the device elements are illustrated in Fig 1b. Settling distances were taken from one element to another element. An overall cross-section area of (20×20) cm$^2$ was designed to fit as a stand-alone air-cleaner or as an air-cleaner for ventilation ducts. This yields a bulk velocity of 2.5 m/s for a flow rate of 100 l/s. However, the bulk velocity increases when the cross-section narrows as in the turbulence-generating grid.

## 2. Methodology

The methodology was divided into three categorises (i) Conventional electrical-mechanical design that led to the concept discussed in Section 1.2, (ii) Aerodynamic study, and (iii) UVC-LED disinfection study. The novelty in this research is within the design of the UVC chamber in terms of the LED layout and the turbulence-generating grid.

*2.2.1 Aerodynamic methodology*

Both one-dimensional (1D) and three-dimensional (3D) analyses were carried out, where the later was based on computational fluid dynamics (CFD). In the 1D analysis, focus was given to the dust filter in front of the air cleaner (see Fig 1), its pressure drop and efficiency.

The pressure-drop $\Delta p$ over the dust-filter was modelled as a netted fibrous filter [17];

$$\Delta p = 64 \mu U_0 h \left( \frac{\alpha_f}{d_f^2} + \frac{\alpha_p}{d_p^2} \right)^{1/2} \left( \frac{\alpha_f}{d_f} + \frac{\alpha_p}{d_p} \right) \quad , \tag{2}$$

where $\alpha_f$ and $\alpha_p$ are the fibre and particle packing relative densities (solidities). $d_f$ and $d_p$ are the cylindrical-fibre and spherical-particle diameters respectively. For our dust-filter $\alpha_f = 0.05$ following a previous study on fibrous filters [18]. When $\alpha_p = 0$ the filter is clean, as $\alpha_p$ increases the filter becomes more clogged and hence $\Delta p$ increases. The pressure-drop $\Delta p$ is linearly proportional to the air dynamic viscosity $\mu$ and the air incoming-speed $U_0$ and thus Stokes flow is assumed around the fibres. *h* is the thickness of the filter

The clean filter efficiency *E* is calculated as [18];

$$E = 1 - e^{\frac{-4 \alpha_p E_T h}{\pi (1 - \alpha_p) d_f}} \quad , \tag{3}$$

The procedure to calculate the total single fibre efficiency (SFE) $E_T$ is given in the appendix. Again, Stokes-flow was assumed around the fibres.

The louvres mounted in front and after the filter also cause a pressure drop. They can be modelled as grids in a similar way as the turbulence-generating grid pressure-drop $\Delta p_g$. The following 1D approximation was derived for a netted grid at a moderate Reynolds number or higher ($Re_{df} > 1000$) which is justified by the centimetre scale of the grid elements [19];

$$\Delta p_g \simeq 0.52 \rho U_0^2 \left[ \frac{1}{(1 - \alpha_g)^2} - 1 \right] \quad , \tag{4}$$

where $\rho$ is the air density, and $\alpha_g$ is the grid solidity. The louvres have solidity much lower than 0.5 and thus their pressure-drop should be less than the dynamic pressure $\rho U_0^2 / 2$.

The skin friction over the device walls and the UVC chamber plates also causes a pressure drop. However, assuming turbulent boundary layers and using the Moody chart, that pressure drop was found to be very small as compared to other contributors, which was confirmed by CFD analysis.

The CFD was based on the RANS approach and the standard k-omega SST model that is suitable for transitional flows. The commercial software Ansys was used along with a second-order upwind scheme for the convective terms and a second-order central scheme for the diffusion terms. No slip wall boundary conditions were applied on the inner walls of the device along standard wall-function for the RANS model. Inflow velocity of 2.5 m/s is imposed for the inlet cross-section area of (20* 20) cm to provide a flow rate of 100 l/s. Unstructured computational mesh was applied along with grid refinement near the grid generating turbulence. Grid convergence was checked for the flow properties discussed in Section 3.

*2.2.2 UVC disinfection methodology*

The inactivation of the aerosolised pathogens in a UVC field with irradiance $S$ (W/m$^2$) can be modelled as [7];

$$\frac{N(t)}{N(t=0)} = e^{-Z \cdot S \cdot t} \quad , \tag{5}$$

where $N$ is the number of active pathogens, $Z$ is the inactivation of constant which is specific for the pathogen and the surrounding medium (m$^2$/J), and t is the residence time (in seconds) of the pathogen in the UVC field. The inactivation constants are given in Table 1. A larger value of $Z$ implies the pathogen disinfection rate is higher.

The irradiance S is calculated using the $\cos^3 \theta$ law that assumes an inverse square distance law, i.e. neglecting scattering and absorption by the air [20]. This is justified by the short distance of radiation of up to 10 cm. Reflections from surfaces that can enhance the irradiance are also neglected and are left for a future study. Considering the LED lamp as a point source mounted on the plate as shown in Fig 1, then at a vertical distance $h_p$ from the plate and at a polar angle $\theta$ from the vertical line;

$$S = \frac{I_0 P(\theta) \cos^3 \theta}{h_p^2} \quad , \tag{6}$$

where $I_0$ can be calculated by taking $S$ = 3.75 mW/cm$^2$ at $h_p$=2 cm just above the LED lamp [20]. The source directivity $P(\theta)$ is [20];

$$P(\theta) = -2 \cdot 10^{-6} \theta^3 - 7 \cdot 10^{-5} \theta^2 + 3 \cdot 10^{-3} \theta + 0.9614 \quad . \tag{7}$$

If $P(\theta) < 0$ then $P(\theta)$ is set to zero. The contributions from different LED lamps are superimposed assuming constructive interaction.

## 3. Results and discussion

The clean dust-filter efficiency $E$ according to Eq (2) is plotted in Fig 3a, showing excellent efficiency for $\alpha_f = 0.05$ and particle size larger than 12 microns. However, one should note that the Reynolds number of the fibre is about 10 and thus Stokes flow does not strictly apply. Nevertheless, Stokes drag law still can give an acceptable estimate in that range [21] and hence the filter with $d_f$ =50 microns was chosen. The filter's pressure drop is shown in Fig 3b, where clogging as expressed by the particle's solidity $\alpha_p$, is seen to significantly increase the pressure drop, hence the importance of an easy replacement of the inexpensive dust-filter as discussed earlier. As the filter will not capture particles smaller than ~12 microns, the particle's clogging can be allowed up to about $\alpha_p = 0.05$ as the chosen fan can overcome up to 200 Pa pressure drop at a power consumption of about 35W.

The dust-filter effectively stops particles large than 12 microns which is more than 90% of the volume of typical household [22]. However, most of the infectious airborne particles are $1\mu m < d_p < 20 \mu m$ [16] and those must be inactivated by the UVC chamber. To achieve sufficient irradiation while having reasonable power consumption, a staggered off-set was applied between the upper and low LED banks as well as between the LED rows in each bank, leading to the novel layout given in Fig 4. The UVC chamber is just 20 cm long and the LEDs rows occupy a spanwise length of 18 cm. The vertical distance between the upper and lower LED banks is 10 cm and thus two irradiation cells in the UVC chamber have an overall cross-section of 20×20 cm. This yields 344 LEDs and a power consumption of 258W.

The irradiance contours as emitted just by the upper LEDs plate and at distances of 4 and 6 cm are shown in Fig 5 with the highest level around the centre as constructive superposition is used between the LEDs. The averaged irradiance per plane in a passage is given in Table 2 where constructive interaction is assumed between the upper and lower LEDs irradiance. This yields a kill of above (97,98,92)% for (SARS-CoV-2, TB, influenza) respectively using the Z values given in Table 1 and assuming a residence time of 0.08 s by taking the length of the chamber divided by the bulk velocity.

The variations in the irradiance levels seen in Fig 5 illustrate the need of well-mixed air for proper irradiance exposure for all particles. This is achieved by the novel turbulence-generating grid (Fig 6), where the variations between the sharp and rounded corners of the large openings as well as the slots increase the turbulence level. As the grid solidity is less than 50%, the estimated pressure drop $\Delta p_g$ = ~9 Pa by Eq (3), where the CFD gave $\Delta p_g$ = ~17 Pa due to the enhanced turbulence. This is still much lower than the pressure drop of the dust filter $\Delta p$ = 150 Pa, assumed for the dust-filter assembly simulation in the CFD computations of Figs 6 and 7. The velocity on the turbulence-generating grid is up to 5.5 m/s with closely uniform distributions inside the grid's holes allowing the sufficient flow rate, as seen in Fig 6b. Typical velocity and turbulent kinetic energy (TKE) fields inside the UVC chamber are shown in Fig 7. High TKE levels are seen in low-velocity zones which will capture and then throw fine particles away from those zones. This is while in other zones the turbulence intensity is between 20% to 10% as typical for fully-developed turbulent channel flow [23], thus having well-mixed air as needed.

## 4. Conclusions

A conceptual design of a UVC-LED air-cleaner based on a dust-filter, a UVC chamber and a fan was introduced. The dust-filter role is to prevent dust accumulation on the LEDs reducing their efficacy. It also reduces the overall particle load in the chamber, that if too high can hamper turbulence levels and minimises potential shading effects due to large particulate matter. However, the filter does not stop the most infectious airborne particles of less than ~12 microns. These are inactivated by the innovative UVC-LEDs layout where a novel turbulence-generating grid is used to enhance the air mixing as shown by CFD results. Irradiance calculations showed a kill of more than 97% for Covid and TB, while maintaining a flow rate of 100 l/s and overall power consumption of less than 300W.

The design can be further improved using reflective paint on the walls to enhance the irradiance. Higher fidelity calculations can be pursued by coupling a radiance transport equation approximation with CFD and particle dynamics while analysing the LEDs heat transfer. The device should also be experimentally tested and further consideration can be given to environmental conditions affecting the pathogen transmissibility [24]. Those steps are to be taken in future planned steps.

**Acknowledgment:**

This study was supported by the UK Royal Academy of Engineering grant EXPP2021\1\247

Table 1: Inactivation constants $Z$ in aerosol [7]

| Pathogen | $Z$ (m$^2$/J) |
|---|---|
| SARS-CoV-2 | 0.377 |
| Mycobacterium Tuberculosis | 0.472 |
| Influenza A | 0.270 |

Table:2 Average irradiance $S$ (mW/cm$^2$) in a single UVC chamber passage of 10 cm height of Fig 5

| Distance from top LED plate (cm) | Irradiance by top LED plate | Irradiance by bottom LED plate | Total irradiance |
|---|---|---|---|
| 1 | 9.82 | 3.52 | 13.34 |
| 2 | 8.38 | 3.98 | 12.36 |
| 3 | 7.34 | 4.48 | 11.82 |
| 4 | 6.47 | 5.06 | 11.53 |
| 5 | 5.72 | 5.72 | 11.44 |
| 6 | 5.06 | 6.47 | 11.53 |
| 7 | 4.48 | 7.34 | 11.82 |
| 8 | 3.98 | 8.38 | 12.36 |
| 9 | 3.52 | 9.82 | 13.34 |

**List of Figures:**

Figure 1: Conceptual design of the UVC-LED air-cleaner in (a) isometric view and (b) longitudinal cross-section where all units are in mm.

Figure 2: Conceptual design of the louvres 'cassette' assembly to provide safe removal of the dust filter. The louvres are closed for the filter's replacement, the all 'cassette' assembly is pulled out from the air-cleaner and the filter can be safely removed.

Figure 3: The dust-filter's (a) efficiency $E$ for several fibre's diameters $d_f$ by Eq (3) and (b) pressure drop $\Delta p$ for $d_f$ = 50 microns and several clogging conditions measured by the particles solidity on the filter $\alpha_p$ using Eq (2), where the filter's solidity is $\alpha_f = 0.05$.

Figure 4: The UVC chamber LEDs layout where (a) staggering is arranged between the upper and lower LEDs arrays and (b) staggering is also shown in the LED array itself to reduce the number of LEDs. The length unit is mm.

Figure 5: The irradiance $S$ (mW/cm$^2$) at a horizontal plane of (20,20) cm as emitted by the upper LED array with the LEDs arrangement of Fig 5b, where the plane is (a) 4 cm and (b) 6 cm away from the upper LEDs array.

Figure 6: (a) The geometry of the turbulence-generating grid where the given dimensions are in mm and the overall dimensions are (200,200) mm (b) The time-averaged velocity magnitude field at the grid.

Figure 7: The (a) time-averaged velocity magnitude and (b) turbulent kinetic energy fields at the mid spanwise plane of the air-cleaner.

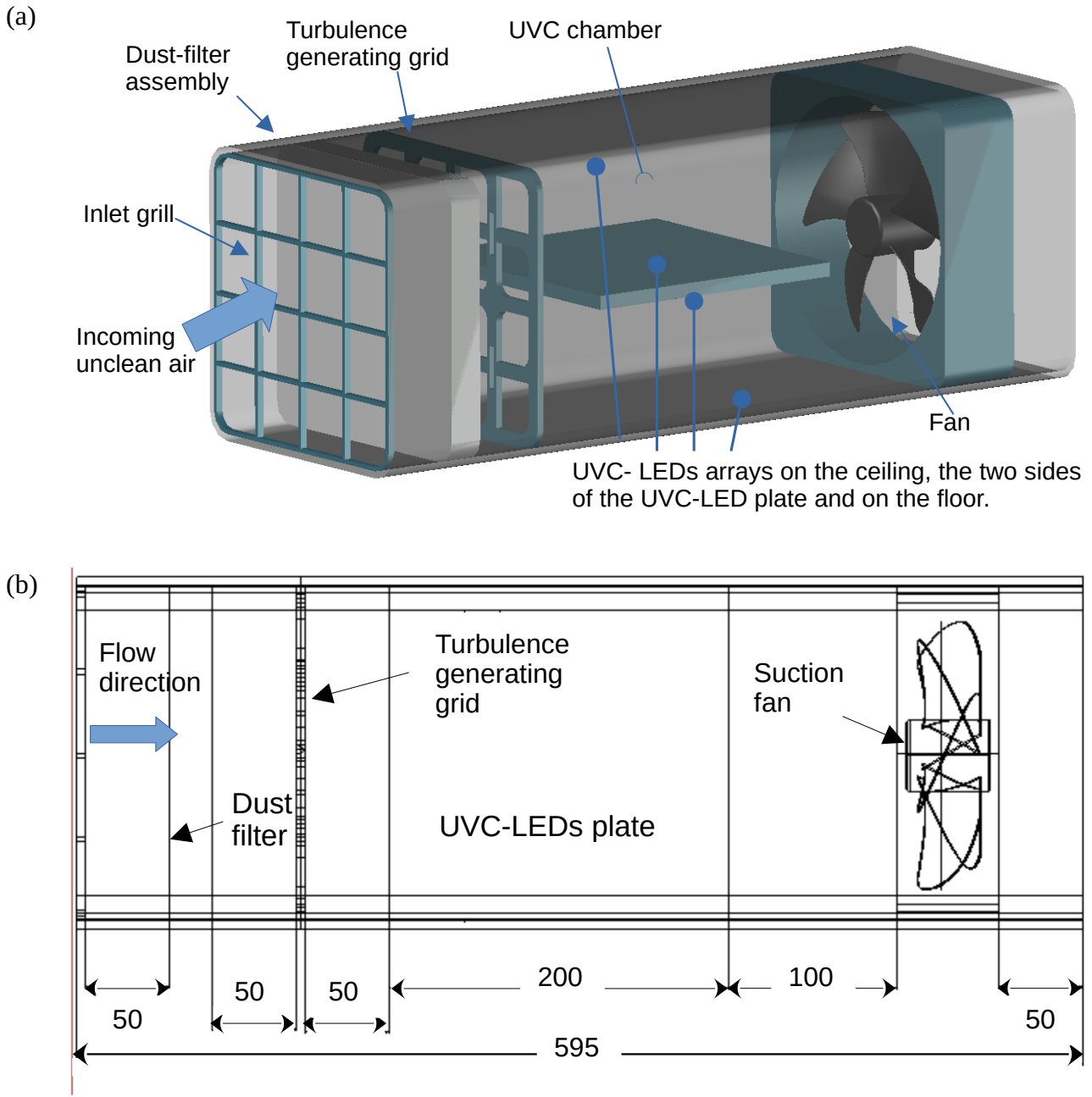

Figure 1: Conceptual design of the UVC-LED air-cleaner in (a) isometric view and (b) longitudinal cross-section where all units are in mm.

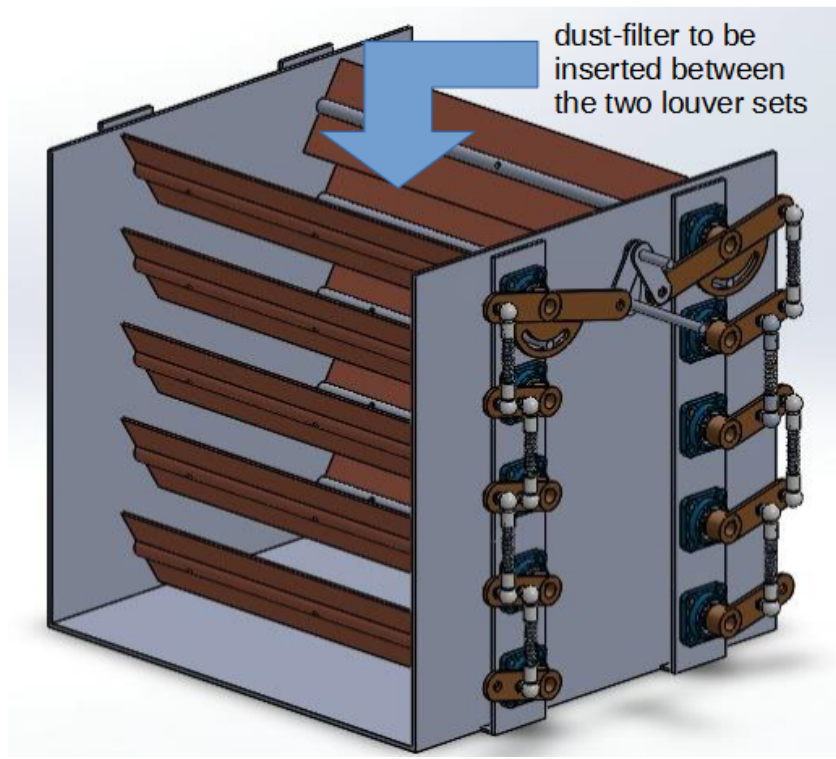

Figure 2: Conceptual design of the louvres 'cassette' assembly to provide safe removal of the dust filter. The louvres are closed for the filter's replacement, the all 'cassette' assembly is pulled out from the air-cleaner and the filter can be safely removed.

(a)

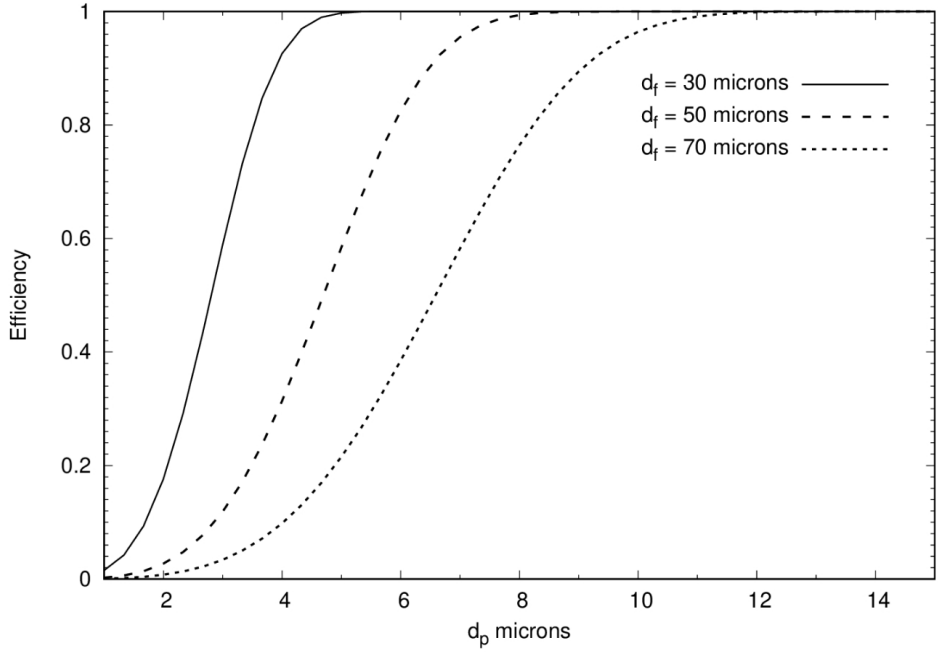

(b)

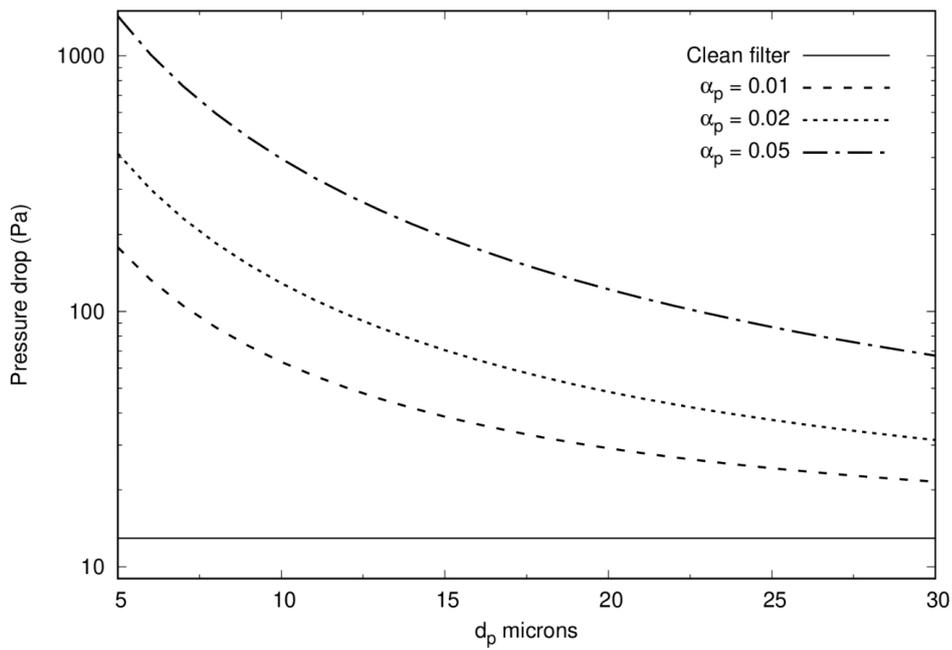

Figure 3: The dust-filter's (a) efficiency $E$ for several fibre's diameters $d_f$ by Eq (3) and (b) pressure drop $\Delta p$ for $d_f$ = 50 microns and several clogging conditions measured by the particles solidity on the filter $\alpha_p$ using Eq (2), where the filter's solidity is $\alpha_f = 0.05$.

(a)

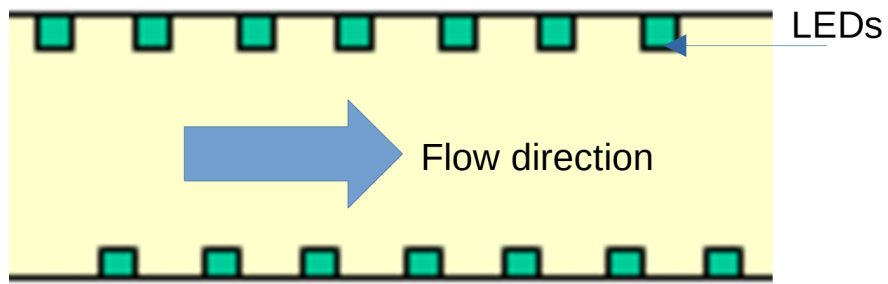

(b)

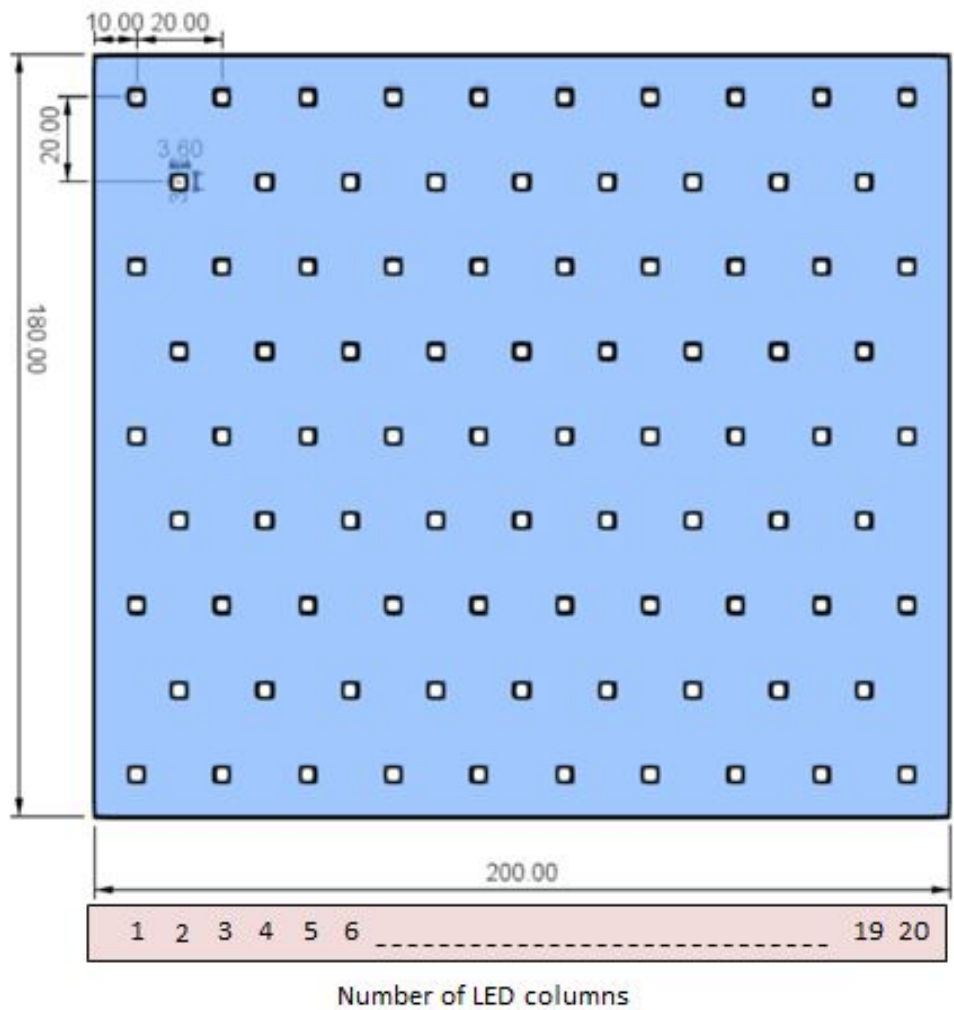

Figure 4: The UVC chamber LEDs layout where (a) staggering is arranged between the upper and lower LEDs arrays and (b) staggering is also shown in the LED array itself to reduce the number of LEDs. The length unit is mm.

(a)

| 2.75 | 3.29 | 3.69 | 4.00 | 4.21 | 4.37 | 4.47 | 4.56 | 4.62 | 4.67 | 4.65 | 4.61 | 4.54 | 4.46 | 4.33 | 4.17 | 3.92 | 3.58 | 3.10 | 2.53 |
|---|---|---|---|---|---|---|---|---|---|---|---|---|---|---|---|---|---|---|---|
| 3.42 | 4.11 | 4.64 | 5.02 | 5.29 | 5.46 | 5.60 | 5.69 | 5.76 | 5.81 | 5.80 | 5.74 | 5.68 | 5.57 | 5.44 | 5.24 | 4.95 | 4.53 | 3.95 | 3.19 |
| 3.97 | 4.83 | 5.48 | 5.94 | 6.26 | 6.48 | 6.63 | 6.75 | 6.82 | 6.89 | 6.88 | 6.82 | 6.73 | 6.62 | 6.46 | 6.23 | 5.90 | 5.41 | 4.73 | 3.86 |
| 4.31 | 5.28 | 6.00 | 6.53 | 6.87 | 7.13 | 7.28 | 7.42 | 7.49 | 7.57 | 7.55 | 7.50 | 7.40 | 7.29 | 7.10 | 6.87 | 6.49 | 5.98 | 5.23 | 4.30 |
| 4.66 | 5.68 | 6.46 | 7.03 | 7.41 | 7.68 | 7.86 | 8.00 | 8.09 | 8.17 | 8.16 | 8.09 | 7.98 | 7.85 | 7.66 | 7.39 | 6.99 | 6.43 | 5.62 | 4.62 |
| 4.85 | 5.88 | 6.69 | 7.27 | 7.68 | 7.95 | 8.16 | 8.29 | 8.40 | 8.47 | 8.47 | 8.38 | 8.29 | 8.14 | 7.94 | 7.65 | 7.25 | 6.65 | 5.83 | 4.77 |
| 5.00 | 6.07 | 6.92 | 7.53 | 7.95 | 8.25 | 8.46 | 8.61 | 8.72 | 8.81 | 8.80 | 8.72 | 8.61 | 8.46 | 8.24 | 7.95 | 7.52 | 6.91 | 6.06 | 4.99 |
| 5.04 | 6.15 | 6.99 | 7.63 | 8.05 | 8.36 | 8.57 | 8.73 | 8.83 | 8.93 | 8.92 | 8.85 | 8.72 | 8.58 | 8.35 | 8.06 | 7.61 | 7.01 | 6.15 | 5.08 |
| 5.18 | 6.28 | 7.15 | 7.79 | 8.23 | 8.55 | 8.77 | 8.93 | 9.05 | 9.14 | 9.13 | 9.05 | 8.92 | 8.76 | 8.53 | 8.22 | 7.77 | 7.14 | 6.27 | 5.17 |
| 5.22 | 6.31 | 7.19 | 7.81 | 8.27 | 8.57 | 8.80 | 8.96 | 9.09 | 9.17 | 9.17 | 9.07 | 8.96 | 8.79 | 8.57 | 8.25 | 7.81 | 7.16 | 6.29 | 5.16 |
| 5.18 | 6.28 | 7.15 | 7.79 | 8.23 | 8.55 | 8.77 | 8.93 | 9.05 | 9.14 | 9.13 | 9.05 | 8.92 | 8.76 | 8.53 | 8.22 | 7.77 | 7.14 | 6.27 | 5.17 |
| 5.04 | 6.15 | 6.99 | 7.63 | 8.05 | 8.36 | 8.57 | 8.73 | 8.83 | 8.93 | 8.92 | 8.85 | 8.72 | 8.58 | 8.35 | 8.06 | 7.61 | 7.01 | 6.15 | 5.08 |
| 5.00 | 6.07 | 6.92 | 7.53 | 7.95 | 8.25 | 8.46 | 8.61 | 8.72 | 8.81 | 8.80 | 8.72 | 8.61 | 8.46 | 8.24 | 7.95 | 7.52 | 6.91 | 6.06 | 4.99 |
| 4.85 | 5.88 | 6.69 | 7.27 | 7.68 | 7.95 | 8.16 | 8.29 | 8.40 | 8.47 | 8.47 | 8.38 | 8.29 | 8.14 | 7.94 | 7.65 | 7.25 | 6.65 | 5.83 | 4.77 |
| 4.66 | 5.68 | 6.46 | 7.03 | 7.41 | 7.68 | 7.86 | 8.00 | 8.09 | 8.17 | 8.16 | 8.09 | 7.98 | 7.85 | 7.66 | 7.39 | 6.99 | 6.43 | 5.62 | 4.62 |
| 4.31 | 5.28 | 6.00 | 6.53 | 6.87 | 7.13 | 7.28 | 7.42 | 7.49 | 7.57 | 7.55 | 7.50 | 7.40 | 7.29 | 7.10 | 6.87 | 6.49 | 5.98 | 5.23 | 4.30 |
| 3.97 | 4.83 | 5.48 | 5.94 | 6.26 | 6.48 | 6.63 | 6.75 | 6.82 | 6.89 | 6.88 | 6.82 | 6.73 | 6.62 | 6.46 | 6.23 | 5.90 | 5.41 | 4.73 | 3.86 |
| 3.42 | 4.11 | 4.64 | 5.02 | 5.29 | 5.46 | 5.60 | 5.69 | 5.76 | 5.81 | 5.80 | 5.74 | 5.68 | 5.57 | 5.44 | 5.24 | 4.95 | 4.53 | 3.95 | 3.19 |
| 2.75 | 3.29 | 3.69 | 4.00 | 4.21 | 4.37 | 4.47 | 4.56 | 4.62 | 4.67 | 4.65 | 4.61 | 4.54 | 4.46 | 4.33 | 4.17 | 3.92 | 3.58 | 3.10 | 2.53 |
| 1.97 | 2.34 | 2.63 | 2.85 | 3.01 | 3.13 | 3.22 | 3.28 | 3.33 | 3.37 | 3.36 | 3.33 | 3.27 | 3.21 | 3.11 | 2.98 | 2.79 | 2.54 | 2.20 | 1.81 |

(b)

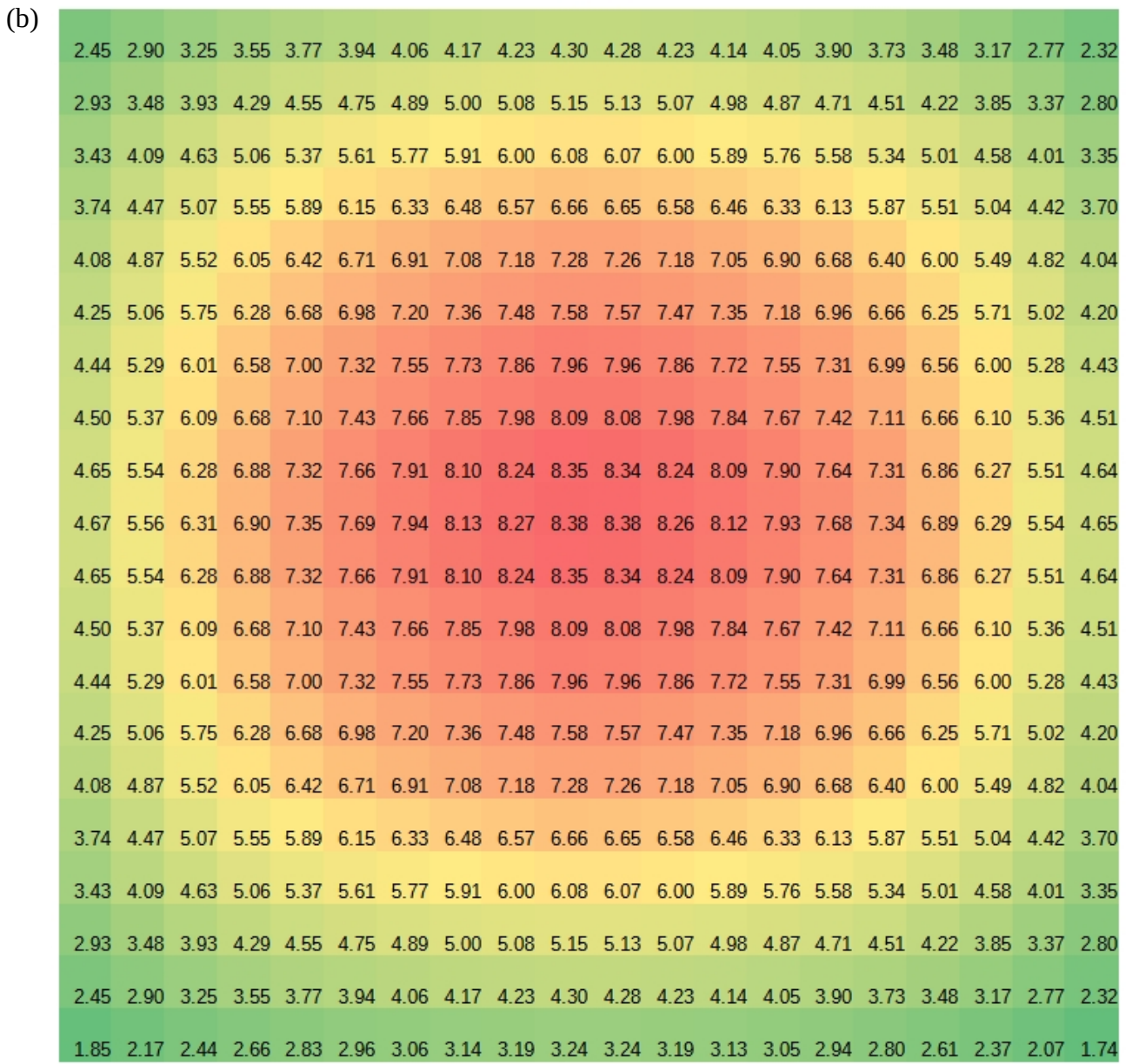

Figure 5: The irradiance $S$ (mW/cm$^2$) at a horizontal plane of (20,20) cm as emitted by the upper LED array with the LEDs arrangement of Fig 5b, where the plane is (a) 4 cm and (b) 6 cm away from the upper LEDs array.

(a)

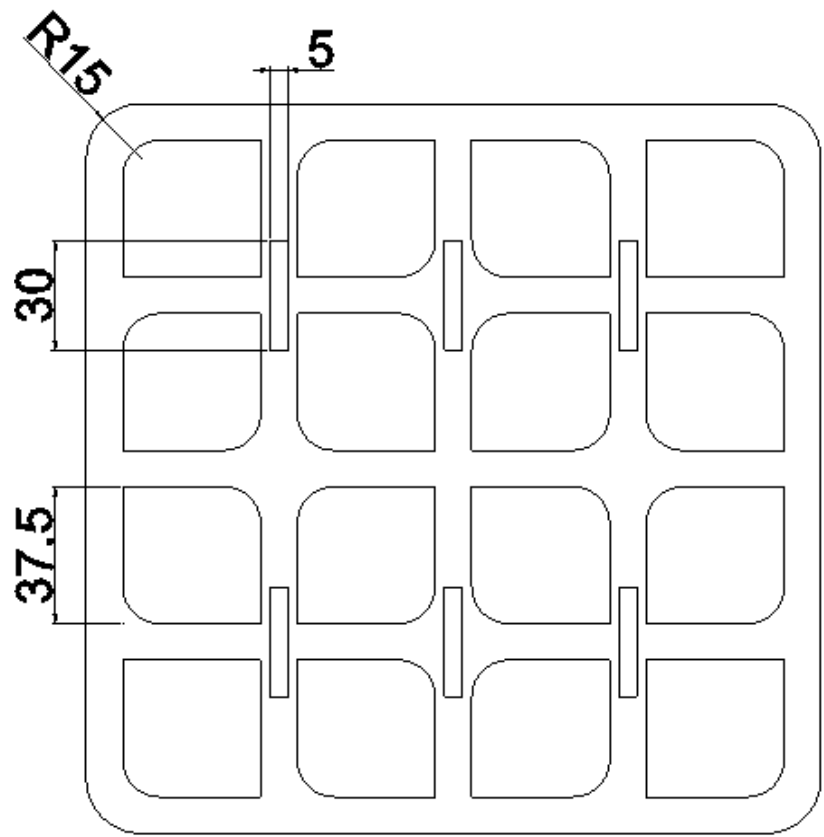

(b)

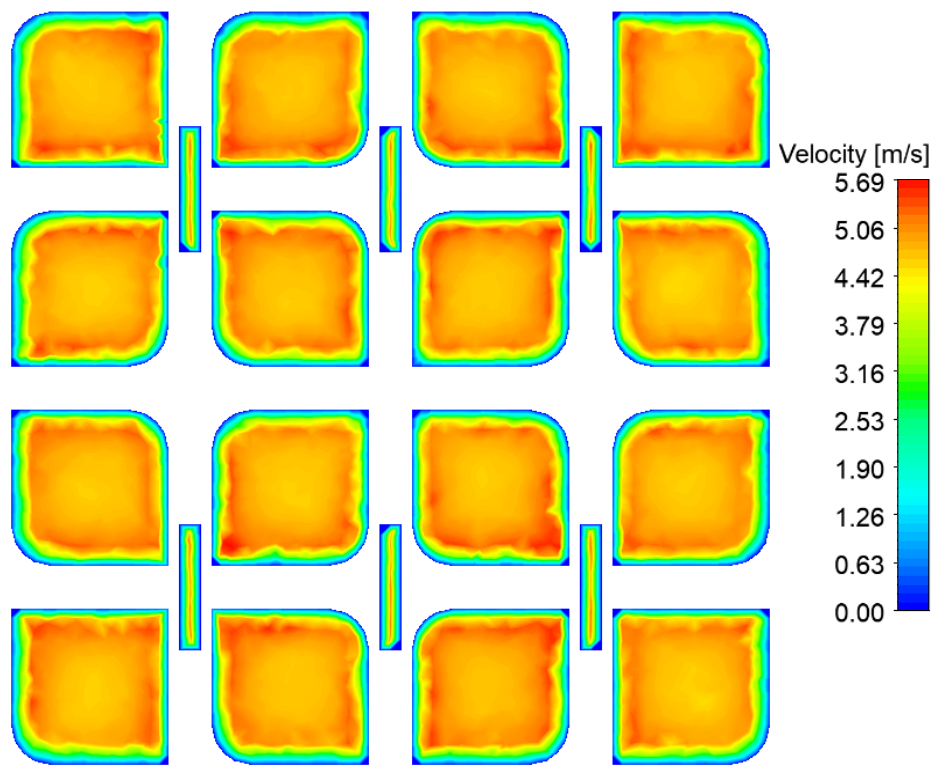

Figure 6: (a) The geometry of the turbulence-generating grid where the given dimensions are in mm and the overall dimensions are (200,200) mm (b) The time-averaged velocity magnitude field at the grid.

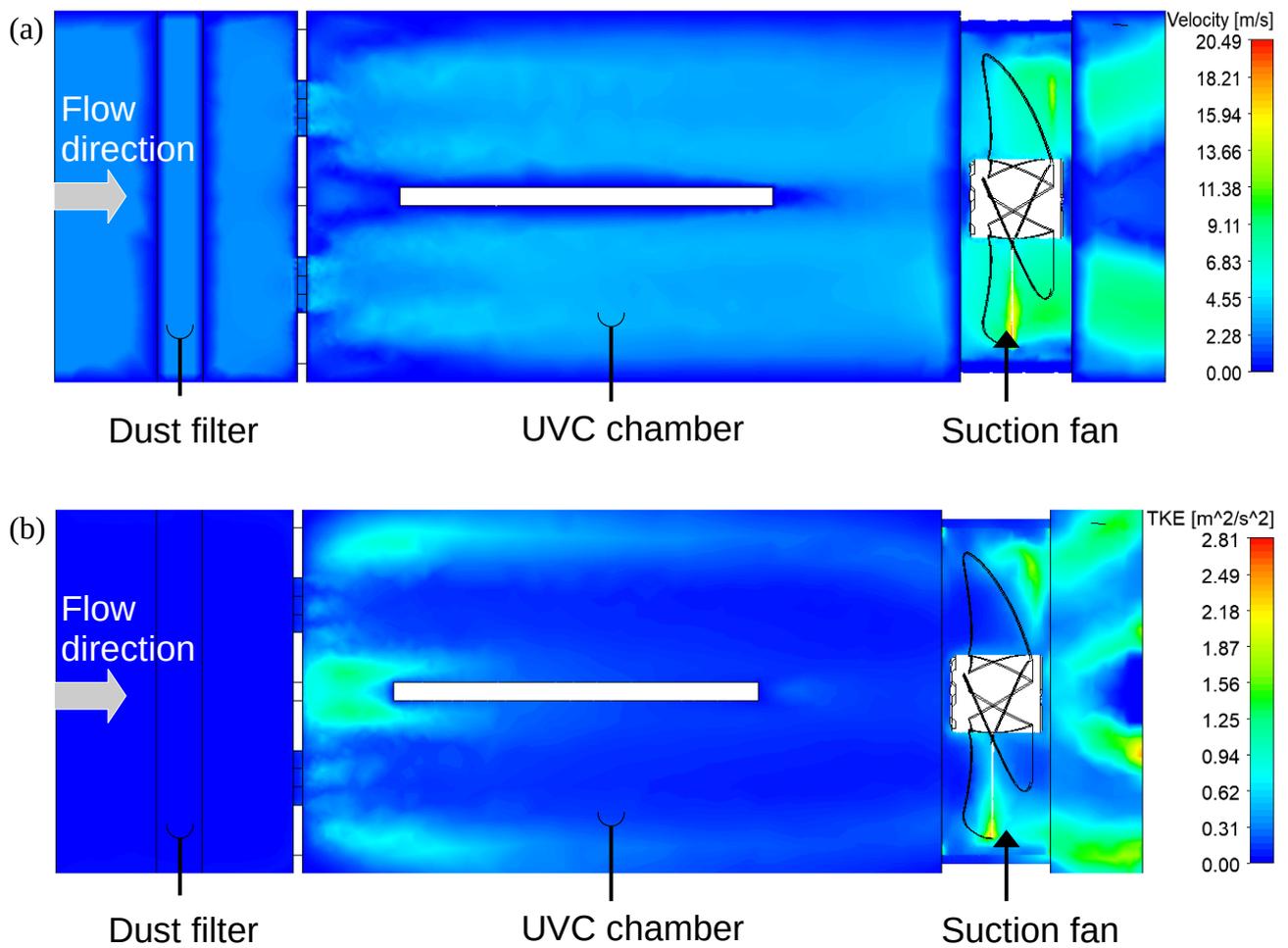

Figure 7: The (a) time-averaged velocity magnitude and (b) turbulent kinetic energy fields at the mid spanwise plane of the air-cleaner.

# Appendix – Dust filter efficiencies

The total single filter efficiency (SFE) $E_T$ is calculated as [17];

$$E_T = 1 - (1-E_D)(1-E_R)(1-E_I) \ . \tag{A1}$$

The diffusion SFE $E_D$ is very small for our filter and thus is neglected. Its calculation procedure can be found in [18]. The interception SFE $E_R$ can be calculated as;

$$E_R = \frac{1-\alpha_f}{K}\left(\frac{R^2}{1+R}\right) \ , \tag{A2}$$

where $\alpha_f$ is the fibre solidity and,

$$K = \alpha_f - \frac{\alpha_f^2}{4} - \frac{\ln \alpha_f}{2} - \frac{3}{4} \ . \tag{A3}$$

$R$ is the ratio between the particle's diameter to the fibre's diameter, i.e. $R = d_p/d_f$. Eq (A3) is considered to be accurate for $R < 0.2$.

The inertial impaction SFE $E_I$ is taken as;

$$E_I = \frac{J \, St}{2K^2} \ , \tag{A4}$$

where

$$J = \begin{cases} (29.6 - 28\alpha_f^{0.62})R^2 - 27.5 R^{2.8} & , R<0.4 \\ 2 & , R>0.4 \end{cases} \ . \tag{A5}$$

The Stokes number $St$ is;

$$St = \frac{d_p^2 \rho_p U_0}{18 d_p \mu} \ , \tag{A6}$$

where the slip factor is taken as one because the particle diameter $d_p$ is of micron-scale and $\rho_p$ is the particle density assumed as of fresh water. Eq (A5) is considered as an accurate approximation when the fibrous filter's solidity is in the range of $0.0035 < \alpha_f < 0.111$, which agrees with our dust filter of $\alpha_f = 0.05$.